\documentclass[symmetry,article,accept,pdftex,moreauthors]{Definitions/mdpi}

\usepackage{rotating}
\usepackage{overpic}
\usepackage{graphicx}
\usepackage{epstopdf}
\usepackage{dcolumn}
\usepackage{bm}
\usepackage{multirow}
\usepackage{slashed}
\usepackage{amssymb,amsfonts}
\usepackage{amsmath}
\usepackage{color}
\def\be{\begin{eqnarray}}
\def\ee{\end{eqnarray}}

\def\lsim{\mathrel{\rlap{\lower3pt\hbox{\hskip1pt$\sim$}}
		\raise1pt\hbox{$<$}}} 
\def\gsim{\mathrel{\rlap{\lower3pt\hbox{\hskip1pt$\sim$}}
		\raise1pt\hbox{$>$}}} 

\firstpage{1} 
\pubvolume{1}
\issuenum{1}
\articlenumber{0}
\pubyear{2024}
\copyrightyear{2024}
\externaleditor{Academic Editor: Firstname \linebreak  Lastname}
\datereceived{22 April 2024 } 
\daterevised{2 June 2024 } 
\dateaccepted{5 June 2024 } 
\datepublished{ } 
\hreflink{\textls[-15]{https://doi.org/}} 

\Title{A Bridge between Trace Anomalies and Deconfinement Phase Transitions}
\TitleCitation{A Bridge between Trace Anomalies and Deconfinement Phase Transitions}

\Author{Bing-Kai Sheng 
 $^{1,2,3}$ and Yong-Liang Ma $^{1,4,}$*}

\AuthorNames{Bing-Kai Sheng and Yong-Liang Ma}

\AuthorCitation{Sheng
, B.; Ma, Y.}

\address{%
$^{1}$ \quad School of Fundamental Physics and Mathematical Sciences,
		Hangzhou Institute for Advanced Study, UCAS, Hangzhou 310024, China; bingkai.sheng@ucas.ac.cn 
\\
$^{2}$ \quad College of Physics, Jilin University, Changchun 130012, China\\
$^{3}$ \quad University of Chinese Academy of Sciences, Beiing 100049, China\\
$^{4}$ \quad School of Frontier Sciences, Nanjing University, Suzhou 215163, China}
\corres{Correspondence: ylma@ucas.ac.cn 
}

\abstract{Inspired by the fact that both the dilaton potential encoding the trace anomalies of QCD and the Polyakov loop potential measuring the deconfinement phase transition can be expressed in the logarithmic forms, as well as the fact that the scale symmetry is expected to be restoring and colors are deconfined in extreme conditions such as high temperatures and/or densities, we conjecture a relation between the dilaton potential and the Polyakov loop potential. Explicitly, we start from the Coleman--Weinberg type potential of a real scalar field---a dilaton or conformal compensator---and make an ansatz of the relation between this scalar field and the Polyakov loop to obtain the Polyakov loop potential, which can be parameterized in Lattice QCD (LQCD) in the pure glue sector. We find that the coefficients of Polyakov potential fitted from Lattice data are automatically satisfied in this ansatz, the locations of deconfinement and scale restoration are locked to each other, and the first-order phase transition can be realized. Extensions to the low-energy effective quark models are also discussed. The conjectured relation may deepen our understanding of the evolution of the universe, the mechanism of electroweak symmetry breaking, the phase diagram of QCD matter, and the properties of neutron~stars.	}

\keyword{Trace anomaly, Polyakov loop potential, deconfinement phase transition 
} 

\begin{document}


\section{Introduction}
It is well known that  massless quantum chromodynamics (QCD) has scale symmetry, but this symmetry is broken by the trace anomaly at the quantum level~\cite{Coleman:1985rnk}. In the low-energy effective field theories (EFTs) of QCD, which are widely used in the study of the non-perturbative phenomena of strong interaction, the trace anomaly effect is implemented through  anomaly matching~\cite{Schechter:1980ak}. Normally, it is written in the form of the Coleman--Weinberg-type logarithmic potential~\cite{Schechter:1980ak,Meissner:1999pe,Goldberger:2007zk,Campbell:2011iw,Matsuzaki:2013eva}.

Although the trace anomaly exists in the QCD vacuum, the scale symmetry is expected to be restored in some extreme conditions, such as high temperatures and/or densities, which may exist in the early universe, heavy-ion collisions at RHIC, LHC, and cores of massive neutron stars. That is, in some extreme conditions, the quantum anomaly effects will be submerged by high temperatures and/or densities due to the vanishming of the beta function of strong  interaction theory. Therefore, the evolution of the scale symmetry, or the physics of dilaton, is interested in cosmology, astrophysics, particle physics, and nuclear physics (see, e.g., Refs.~\cite{Gasperini:2007ar,Sasaki:2011ff,Crewther:2013vea,Ma:2020tsj,Fujimoto:2022ohj} and references therein).

In these extreme conditions, quarks are deconfined~\cite{Fukushima:2010bq,Fukushima:2011jc,Petreczky:2012rq,Adams:2012th,Andersen:2021lnk,Kharzeev:2015kna,Andersen:2014xxa}, and this deconfinement phase transition is mimicked in effective models through a Polyakov loop $\Phi$~\cite{Fukushima:2003fw}, which is defined as the trace of the Wilson line in Euclidian space in the temporal direction~\cite{Ratti:2021ubw}. Physically, the expectation value $\left\langle\Phi \right\rangle $ is capable being of the order parameter of the deconfinement phase transition since $\left\langle\Phi\right\rangle \propto e^{-(F-F_{0})/T}$ in the heavy-quark limit, where $F_{0}$ is the free-energy of gluons and $F-F_{0}$ means the least work to excite a quark in the thermal gluon medium. In terms of this relation, it is obvious that when $\left\langle\Phi \right\rangle=0$, the system is in the confined phase with an infinite amount of energy to pull out a quark from the system, but when $\left\langle\Phi \right\rangle\neq 0$, it is in the deconfined phase, and states with a single quark are possible~\cite{Ratti:2021ubw}.

As an ab initio calculation, Lattice QCD (LQCD) can address the issues related to the deconfinement phase transition and the restoration of scale symmetry at high temperatures and therefore provides abundant phenomena that can help us deepen our understanding of the strong interaction. However, finite quark chemical potential disables the Monte Carlo simulations when  the numerical path integral is performed on a large discretized Euclidian space--time lattice since the exponential of the QCD action is no longer a real number and is not a weight function for the field configurations. This is known as the fermion sign problem~\cite{Dexheimer:2020zzs,Troyer:2004ge}. For more details about LQCD, refer to, e.g., Refs.~\cite{MUSES:2023hyz,Ratti:2021ubw} and references therein. The behavior of the Polyakov loop as a function of temperature is calculated on the lattice for a pure glue system~\cite{Kaczmarek:2002mc,Ratti:2005jh},  the deconfinement phase transition is of the first order, and the transition temperature is approximately $270$~MeV. For a system involving $2+1$ dynamical quark flavors, the lattice calculation indicates that the transition is a smooth crossover~\cite{Ratti:2010kj}, and the Polyakov loop is no longer an order parameter due to the explicit $\mathbb{Z}_{3}$ symmetry breaking by quarks~\cite{Ratti:2021ubw}.

Generally, the Polyakov loop potential is approximated as the polynomial of the Polyakov loop $\Phi$, and its conjugation with the parameters is fixed by fitting the Lattice data~\cite{Pisarski:2000eq,Ratti:2005jh,Pisarski:2006hz,Schaefer:2007pw}. However, a simplified version of the general form of the Polyakov loop potential consists of a logarithmic function of $\Phi$ in addition to a $\Phi^2$ term~\cite{Fukushima:2003fw,Roessner:2006xn,Dexheimer:2009hi}.

Considering that the scale symmetry will be restored and the colors are deconfined in the extreme conditions, we conjectured a lock between the elimination of the trace anomaly and the deconfinement transition or a relation between the dilaton field $\chi$ (or so-called conformal compensator) and Polyakov loop $\Phi$ based on the fact that both $\chi$ and $\Phi$ fields are related to the gluonic configuration according to the anomaly matching~\cite{Schechter:1980ak} and the definition of the Polyakov loop~\cite{Ratti:2021ubw}. Explicitly, regarding $\left\langle \chi\right\rangle$ as a function of  $\left\langle\Phi\right\rangle$ in the sense of the mean field approximation (MFA), and considering that $\left\langle \chi\right\rangle(\left\langle\Phi\right\rangle=0)\neq 0$ means the trace anomaly exists in the confined phase and $\left\langle \chi\right\rangle(\left\langle\Phi\right\rangle=1)=0$ indicates the restoration of the scale symmetry in the deconfined phase, we introduce a function of $ \left\langle \chi\right\rangle$, which is a polynomial of $\left\langle\Phi\right\rangle$ and $\left\langle\Phi\right\rangle^{\ast}$ up to $\mathcal{O}(\left\langle\Phi\right\rangle^4)$ and $\mathbb{Z}_{3}$ is symmetric. It is extremely interesting that the coefficients of the Polyakov potential fitted from Lattice QCD (LQCD) automatically satisfy the condition $\left\langle \chi\right\rangle(\left\langle\Phi\right\rangle=1)=0$. The logarithmic Polyakov potential, which is similar to that in Refs.~\cite{Dexheimer:2009hi,Dexheimer:2008av,Dexheimer:2020rlp} can be yielded after substituting $\chi(\Phi,\Phi^{*})$ into the Coleman--Weinberg type potential $V(\chi)$.

The relation suggested in this work sets up a bridge between the restoration of scale symmetry and deconfined phase transition; therefore, it may deepen our understanding of the evolution of the universe, the mechanism of the electroweak symmetry breaking, the phase diagram of QCD, and the properties of compact stars and gravitational waves.

\section{Trace Anomaly and the Polyakov Loop}	
In the effective models in terms of hadrons, the trace anomaly of QCD can be described effectively by the Coleman--Weinberg type potential in the form~\cite{Schechter:1980ak,Meissner:1999pe,Goldberger:2007zk,Campbell:2011iw,Matsuzaki:2013eva,Papazoglou:1998vr}
\be
V(\chi)=\dfrac{m_{\sigma}^{2}f_{\sigma}^{2}}{4}\left(\dfrac{\chi}{f_{\sigma}}\right)^{4}\left[\mathrm{ln}\left(\dfrac{\chi}{f_{\sigma}}\right)-\dfrac{1}{4}\right],
\label{eq:dilatonPotential}
\ee	
where the field $\chi$ is the conformal compensator, $m_\sigma$ is the dilaton mass with $f_\sigma$ as its decay constant. Potential \eqref{eq:dilatonPotential} is widely used in particle physics and astrophysics to understand the properties of Higgs and the compact star matter (see, e.g., Refs.~\cite{Campbell:2011iw,Yamawaki:2015tmu,Ma:2019ery} and references therein).

\begin{figure}[H]

\includegraphics[scale=0.08]{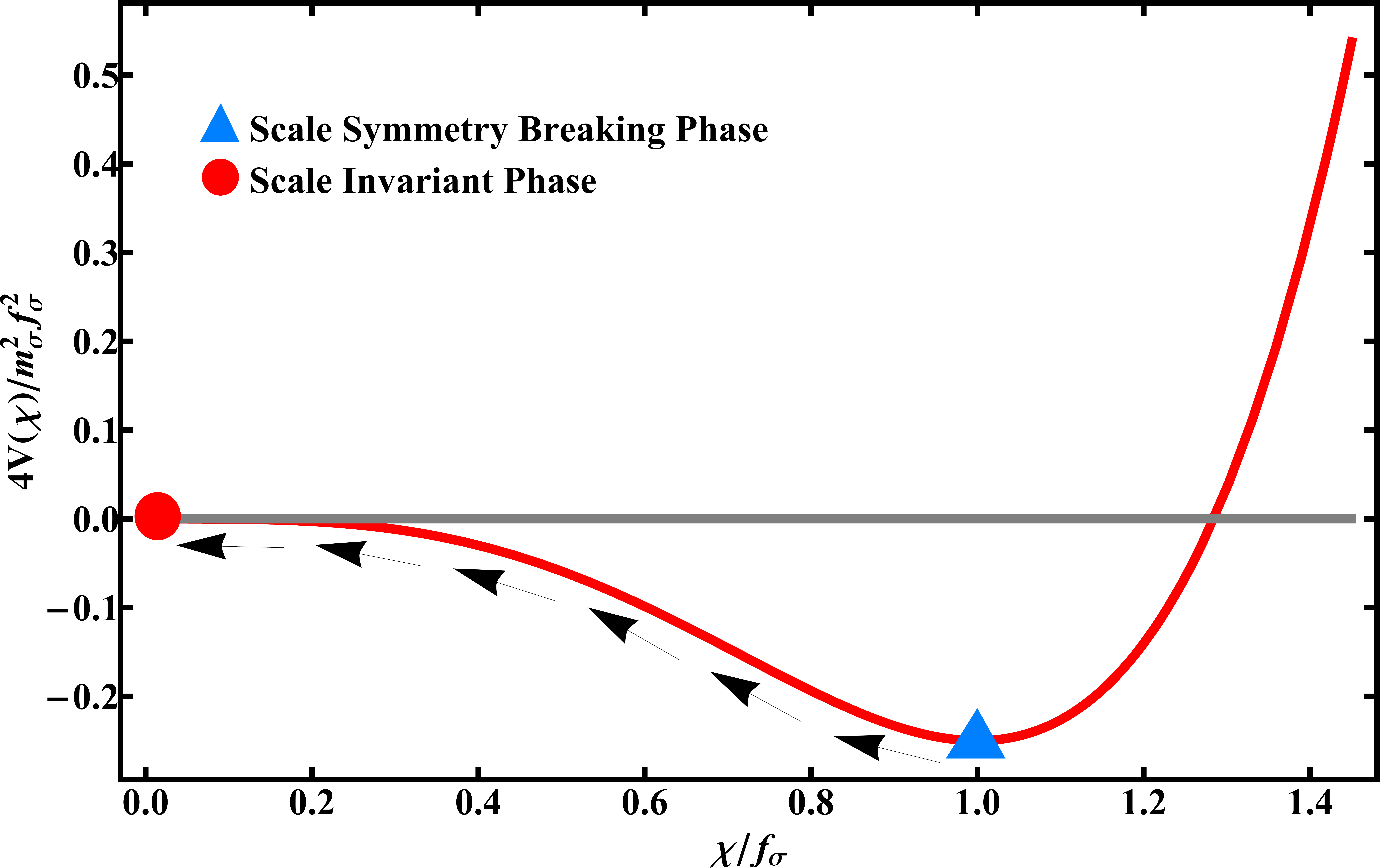}	
\caption{The profile of the dilaton potential. The arrows indicate the evolution of the potential with the increase in temperature and/or density.}
\label{Fig:dilatonPotential}
\end{figure}

From the profile of the dilaton potential sketched in Figure~\ref{Fig:dilatonPotential}, one can easily see that there is a nontrivial vacuum at $\chi=f_{\sigma}$, which breaks the scale symmetry spontaneously. When the temperature and/or density increase, this nontrivial vacuum will approach the zero point $\chi=0$ and the scale symmetry is restored by the extreme environments.

On the other hand, in the framework of statistical mechanics, the deconfinement phase transition can be described effectively by the Polyakov loop~\cite{Polyakov:1978vu}
\be
\Phi(\bm{x})=\frac{1}{N_{c}}\mathrm{tr_{c}}\left[\mathcal{P}\exp\left[-ig_{s}\int_{0}^{\beta}\mathrm{d}x_{4}A_{4}^{a}(\bm{x},x_{4})T^{a}\right]\right],
\label{eq:Polyakovloop}
\ee
with $N_{c}=3$ and $A_{4}^{a}(\bm{x},x_{4})$ being the temporal component of the gluon field, $\mathcal{P}$ being the path-ordering operator, and $T^{a},a=1,\cdots,8$ being the generators of $SU(3)$ gauge group. In the confined phase, $\left\langle\Phi\right\rangle=0$ and, with the increasing of temperature and/or density, it monotonically increases to $\left\langle\Phi\right\rangle=1$ in the deconfined phase. In terms of $\Phi$, the Polyakov potential  generally has two versions in which  one is the simplest polynomial form consisting of terms $\Phi^{*}\Phi$, $\Phi^{*3}+\Phi^{3}$ and $(\Phi^{*}\Phi)^{2}$, which are $\mathbb{Z}_{3}$ symmetric---see, e.g., Ref.~\cite{Ratti:2005jh}---and the other is an improved expression of which the higher order polynomial terms are replaced by the logarithm of a $\mathbb{Z}_{3}$ symmetric function $J(\Phi,\Phi^{*})$~\cite{Fukushima:2003fw,Roessner:2006xn}. These Polyakov potentials contain no baryon density effect due to the lack of the baryon chemical potential. The authors of Ref.~\cite{Dexheimer:2009hi} proposed a simplified version of the potentials used in~\cite{Ratti:2005jh,Roessner:2006xn}, and they incorporated terms that are dependent on the chemical potential to address the issues of compact stars. At zero baryon density, this kind of Polyakov potential reads
\be
\mathcal{U}(\Phi) & = & a_{3}T_{0}^{4}\mathrm{ln}\left(1-6\Phi^2+8\Phi^3-3\Phi^4\right) + a_{0}T^{4}\Phi^2.
\label{eq:PolyakovPotential}
\ee
Here, the transition temperature of the pure glue system $T_{0}$ is fixed to $270$~MeV~\cite{Fukushima:2003fw,Ratti:2005jh,Roessner:2006xn,Schaefer:2007pw,Dexheimer:2009hi}, and $a_{0}=-1.85, a_{3}=-0.40$, which are determined by reproducing the Lattice data and the information about the phase diagram~\cite{Dexheimer:2009hi}.

It is interesting to note that both the dilaton potential~\eqref{eq:dilatonPotential} and the Polyakov potential~\eqref{eq:PolyakovPotential} are logarithmic. Inspired by this similarity, with respect to the picture in which the trace anomaly   vanishes and colors are deconfined at high temperatures and/or densities and the fact that both $\chi$ and $\Phi$ depend on the configuration of the gluon field, we make a conjecture of the $\mathbb{Z}_{3}$ symmetric relation between the dilaton field and the Polyakov loop in MFA as follows:
\be
\dfrac{\left\langle\chi\right\rangle}{f_{\sigma}} & = & 1-a\left\langle\Phi\right\rangle^{*}\left\langle\Phi\right\rangle - \dfrac{b}{2}\left(\left\langle\Phi\right\rangle^{*3}+\left\langle\Phi\right\rangle^{3}\right) \nonumber \\
& &{} - c\left(\left\langle\Phi\right\rangle^{*}\left\langle\Phi\right\rangle\right)^{2}.
\label{eq:mfa}
\ee
When the deconfinement phase transition happens, the field $\chi$ will vanish, which means that the degree of freedom of the dilaton will disappear with the emergence of the gluonic degree of freedom and the potential $V$ will also vanish corresponding to the restoration of the scale symmetry. Hence, we have $\left\langle\chi\right\rangle(\left\langle\Phi\right\rangle=1)=0$, namely,
\be
a+b+c=1.
\label{eq:constraintabc}
\ee

Note that since the physical vacuum depends on the temperature, namely the expectation value of the Polyakov loop is a function of the temperature, the values of the parameters $a$, $b$, and $c$ should be temperature-modified quantities, like what happens in the dense system where the parameters in the model are density dependent ones through the well-known Brown--Rho scaling~\cite{Brown:1991kk}. Like in the dense system, we call this temperature dependence  an intrinsic temperature dependence (ITD), and the temperature effect in the physical system should include both ITD and the effect of to hadron fluctuations (see, e.g.,~\cite{Ma:2019ery})
.

The constraint~\eqref{eq:constraintabc} is fixed at the deconfinement temperature. We simply assume that the temperature dependences of the parameters satiate this constraint. We admit that we cannot explicitly fix these temperature dependences from fundamental QCD. One might give an explicit dependence using the holographic QCD model in which the temperature effect enters through the Hawking temperature of a black hole in the AdS bulk, i.e., the five-dimensional anti-de Sitter space--time.

Substituting Equation~\eqref{eq:mfa} into Equation~\eqref{eq:dilatonPotential}, one can express the dilaton potential in terms of the Polyakov loop as $V(\Phi,\Phi^\ast)$, therefore connecting the trace anomaly with quark confinement. If the baryon or quark density is zero, the Polyakov loop will be a real field~\cite{Ishii:2015ira}, and the dilaton potential will read
\be
V(\Phi)&=&{\dfrac{m_{\sigma}^{2}f_{\sigma}^{2}}{4}\left(1-a\Phi^{2}-b\Phi^{3}-c\Phi^{4}\right)^{4}\left[\ln\left(1-a\Phi^{2}-b\Phi^{3}-c\Phi^{4}\right)-\dfrac{1}{4}\right]} \nonumber \\
&=&\dfrac{m_{\sigma}^{2}f_{\sigma}^{2}}{4}\mathrm{ln}\left(1-a\Phi^{2}-b\Phi^3-c\Phi^4\right)+\cdots.
\label{eq:DilatonPolyakovR}
\ee
We can expand the factor $(1-a\Phi^{2}-b\Phi^{3}-c\Phi^{4})^{4}$ (the primary contribution of $(1-a\Phi^{2}-b\Phi^{3}-c\Phi^{4})^{4}$ is the lowest order of $\Phi$ due to $\Phi<1$)  in the first line of the above equation, and the lowest order of $\Phi$ multiplied by the logarithmic term is shown explicitly in the second line. Comparing Equations~(\ref{eq:PolyakovPotential}) and (\ref{eq:DilatonPolyakovR}), one finds that a $\Phi^2$ term should be added to the potential $V(\Phi)$, and therefore, the Polyakov potential incorporated the trace anomaly is as follows~(it should be noted that the term $\frac{m_{\sigma}^{2}f_{\sigma}^{2}a}{16}\Phi^2$ exists in the straightforward expansion of $V(\Phi,\Phi^\ast)$, but with only the potential V, the first-order phase transition in the pure glue theory cannot be realized):
\be
\mathcal{U}(\Phi,\Phi^{*})=V(\Phi,\Phi^{*})+\dfrac{m^{2}_{\sigma}f^{2}_{\sigma}}{4}d\Phi^{*}\Phi,
\label{eq:FinalPotent}
\ee
with $d$ being a new parameter. It should be noted that in the deconfinement phase ,the potential of the trace anomaly must vanish, i.e., $V(\left\langle\Phi\right\rangle=1)=0$, but $\mathcal{U}(\left\langle\Phi\right\rangle=1)=\dfrac{m^2_{\sigma}f^2_{\sigma}}{4}d\neq 0$.

Comparing~\eqref{eq:PolyakovPotential} and~\eqref{eq:DilatonPolyakovR}, one can relate the transition temperature to the dilaton properties as follows:
\be
T_{0} & \equiv & \left(\dfrac{m_{\sigma}^2f_{\sigma}^2}{4|a_{3}|}\right)^{1/4} \approx 275~\text{MeV},
\label{eq:T0}
\ee
where $m_\sigma=640~$MeV and $f_\sigma=150~$MeV which are consistent with those used in the chiral-scale effective theory~\cite{Crewther:2013vea,Li:2016uzn,Ma:2016nki,Shao:2022njr}. Interestingly, the present naive estimation of $T_{0}$ is close to that from LQCD ($T_{c}=270~\text{MeV}$), and it is amazing that the coefficients of $\Phi^{2}$, $\Phi^{3}$ and $\Phi^{4}$ in the logarithmic term (i.e., $a=6, b=-8, c=3$) of the potential Equation~\eqref{eq:PolyakovPotential} obtained from LQCD satisfy the constraint \eqref{eq:constraintabc}, which implies the relation of the $\chi$ field and $\Phi$ field Equation~\eqref{eq:mfa} is reasonable.
If our ansatz makes sense, we may utilize our Polyakov potential Equation~\eqref{eq:FinalPotent} to investigate the deconfinement transition such as the order of the transition and the equation of state of dense nuclear matter based on low-energy models of QCD. 
	
\section{Qualitative Analysis}
We next analyze the potential~\eqref{eq:FinalPotent} in the thermal system more closely with respect to the data from LQCD simulations~\cite{Kaczmarek:2002mc,Ratti:2005jh}. 

There are four parameters, $a, b, c$ and, $d$, in~\eqref{eq:FinalPotent}, and three of them, like  $a,b$, and $d$, can be taken as independent ones due to the constraint~\eqref{eq:constraintabc}. We find that $a$ and $d$ can be fixed as $a=6,d=-0.1$, and $b$ is a function of the temperature such that the first-order phase transition can be realized. 

\begin{figure}[H]	

\includegraphics[scale=0.06]{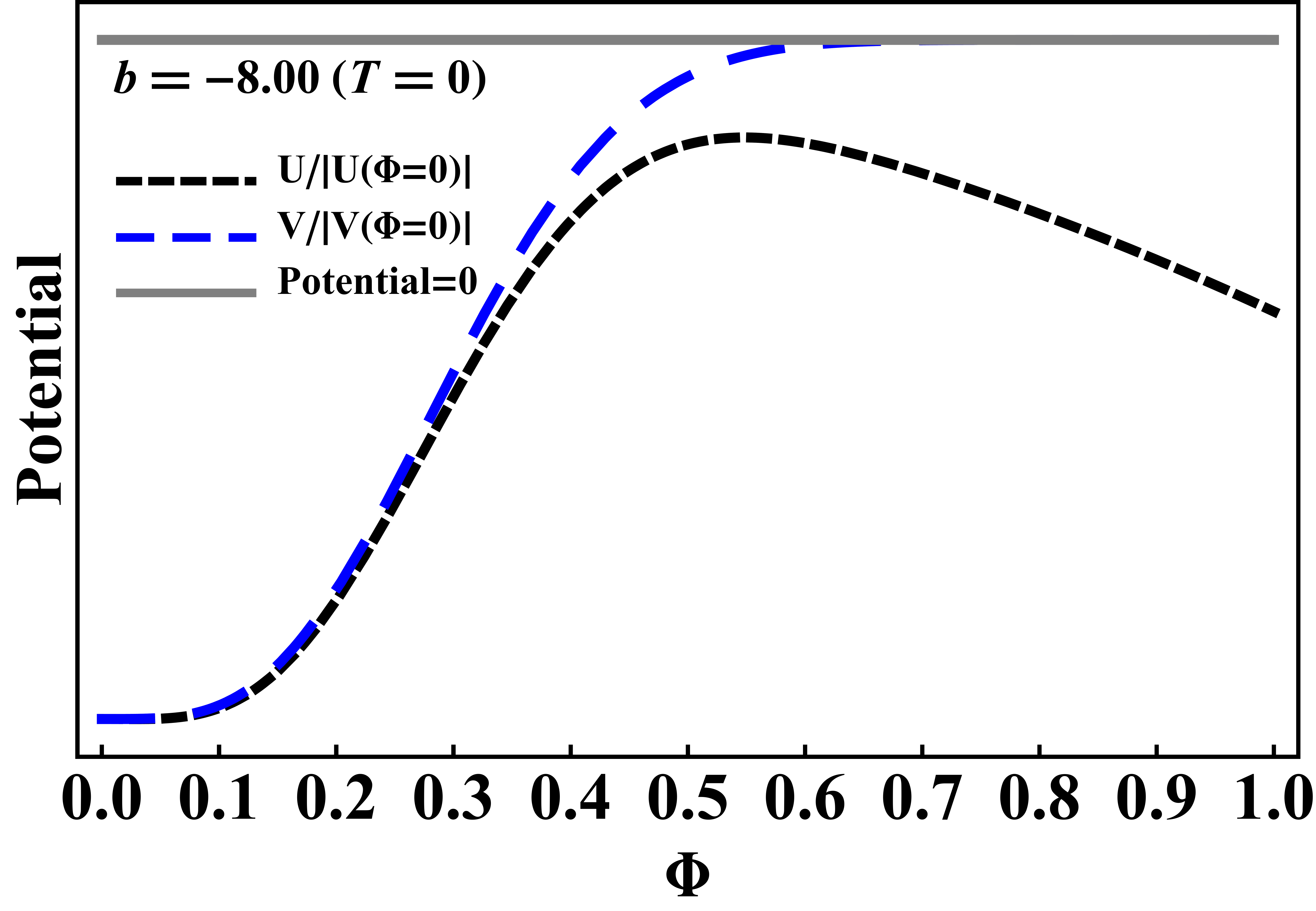}\quad	\includegraphics[scale=0.06]{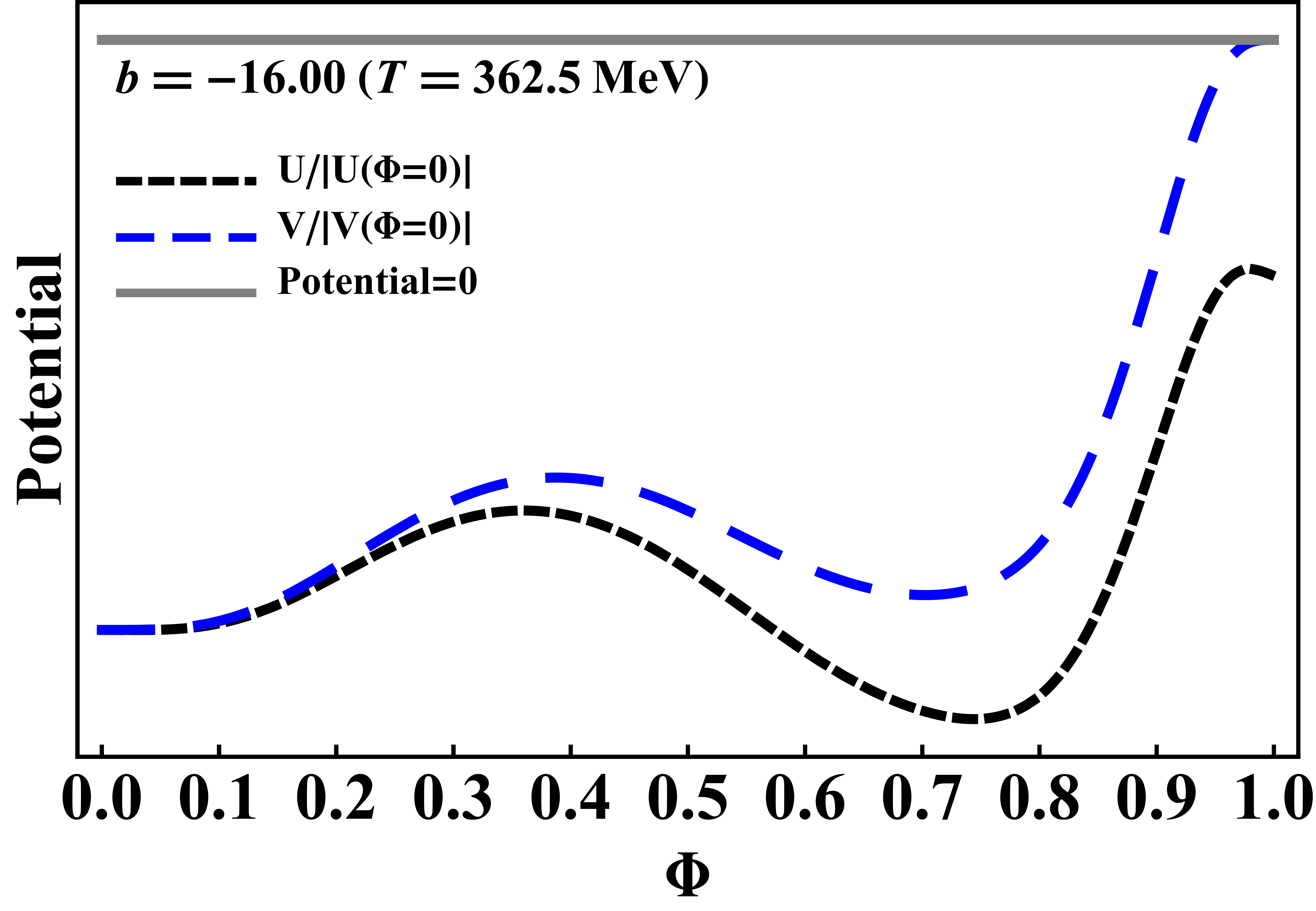}\quad
\includegraphics[scale=0.06]{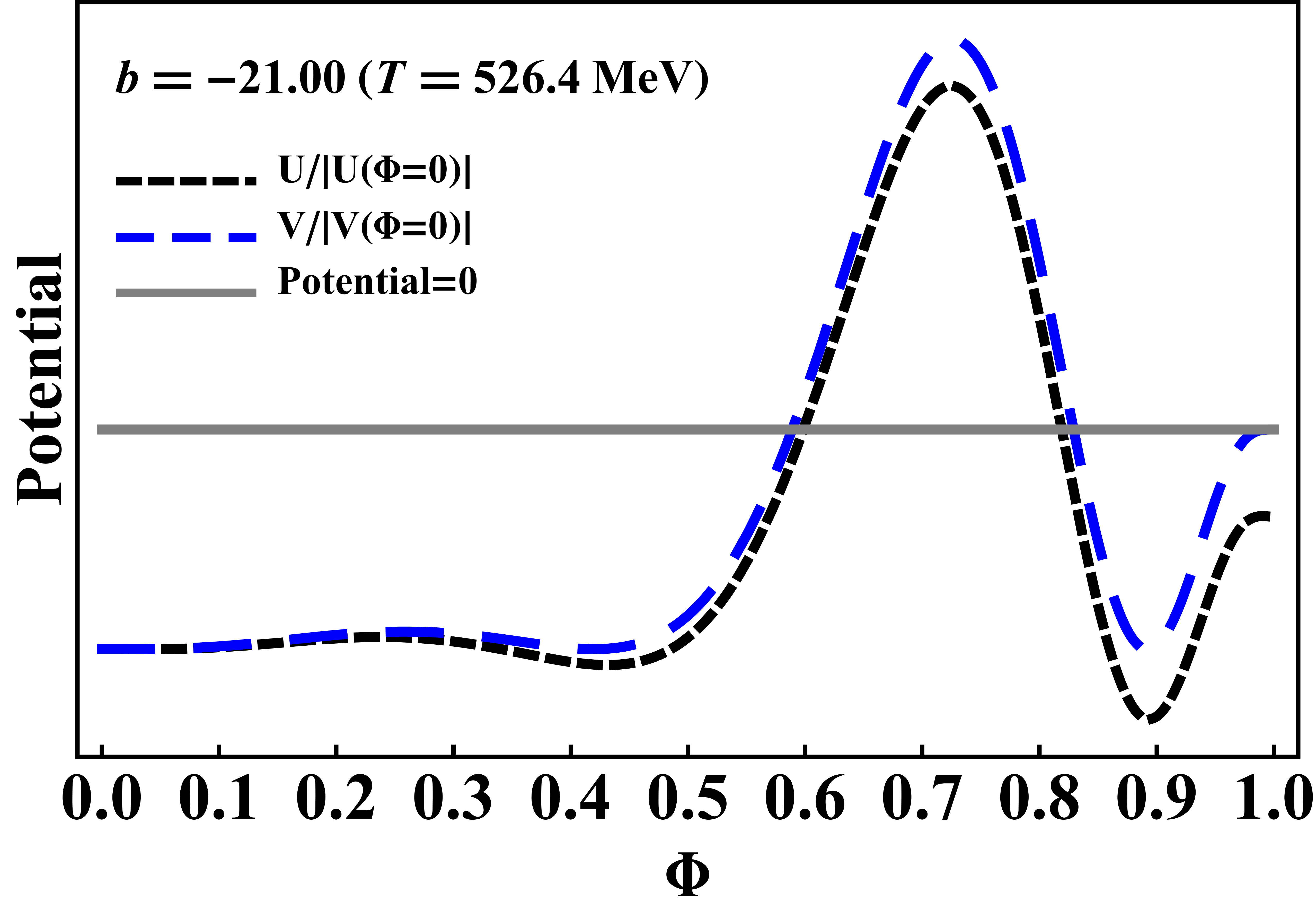}\quad	\includegraphics[scale=0.06]{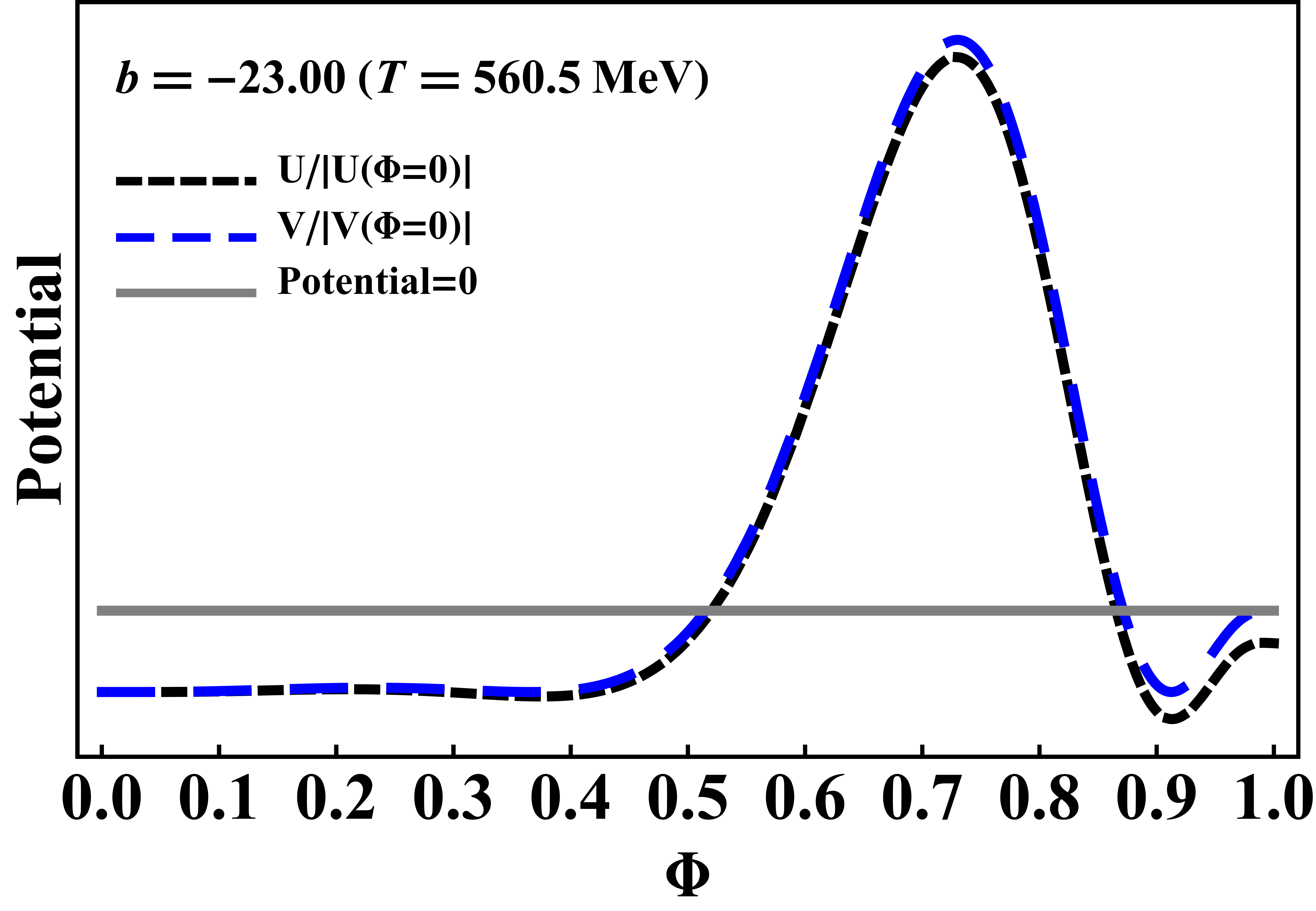}	
\caption{The curves of $\mathcal{U}(\Phi)$ and $V(\Phi)$ as the functions of $\Phi$. From left to right, the temperature is increasing from zero, with $b(T=0)=-8$, $b(T=362.5~\text{MeV})=-16$, $b(T=526.4~\text{MeV})=-21$, and $b(T=560.5~\text{MeV})=-23$ fixed by using LQCD data~\cite{Kaczmarek:2002mc,Ratti:2005jh}, i.e., the magnitude of $b$ is increasing with the growth of temperature. The gray line corresponding to the zero of the potential is shown for the sake of clarity.}
\label{fig:VPhi}
\end{figure}

In Figure~\ref{fig:VPhi}, we sketch the Polyakov potentials at different temperatures---different values of $b$. The first figure shows that $\left\langle \Phi\right\rangle=0$ and the system is in the confinement phase at low temperature. With the increasing temperature, another local minimum point emerges, and at the critical temperature $\widetilde{T}_{c}$, the vacuum will jump from the location of $\left\langle\Phi\right\rangle=0$ to that of $\left\langle\Phi\right\rangle\neq 0$, i.e., $\left\langle\Phi\right\rangle\approx 0.75$, and thus the first-order phase transition will take place as shown in the second figure. The last two figures show that on the one hand, the equality of minimal values of $V$ remains unchanged with the increase in temperature, and this issue implies that the Polyakov potential contains not only the contribution of the trace anomaly but also the extra quadratic term of $\Phi$, which can pull the minimum point of $\left\langle\Phi\right\rangle\neq 0$ down so that the first-order phase transition can be realized; on the other hand, the minimum point of $V$ will approach the gray line when the temperature increases, implying that the trace anomaly will vanish at the high-temperature limit. 

\begin{figure}[H]

	\includegraphics[scale=0.1]{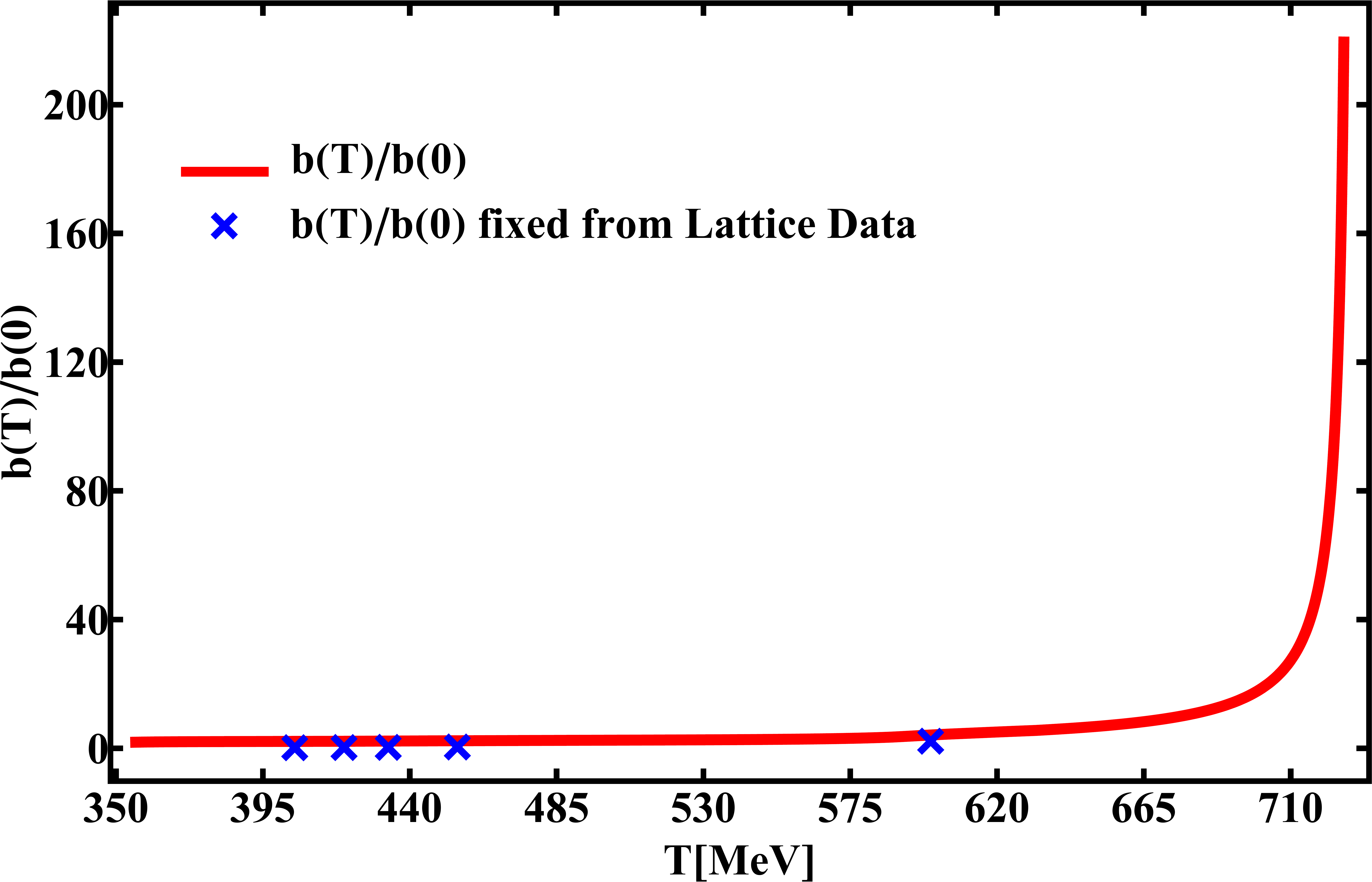}	
	\caption{\label{fig:b_T}The parameter $b(T)$ as a function of temperature fixed in terms of the $\left\langle\Phi\right\rangle(T)$ calculated by LQCD, here $b(0)=-8$.}
\end{figure}

In Figure~\ref{fig:b_T}, we estimate the parameter $b(T)$ as a function of temperature, utilizing the $\left\langle\Phi\right\rangle$ from LQCD~\cite{Kaczmarek:2002mc,Ratti:2005jh}. Specifically, we first obtain the function $\left\langle\Phi\right\rangle(b)$ according to the condition $\left.\frac{\partial \mathcal{U}}{\partial \Phi}\right|_{\Phi=\left\langle\Phi\right\rangle}=0$ and then extract the $\left\langle\Phi\right\rangle(T)$ from the data of LQCD. We then obtain the function $b(T)$ numerically by canceling the variable $\left\langle\Phi\right\rangle$. It should be emphasized that since the phase transition is first-order and $\left\langle\Phi\right\rangle $ is always zero at low temperatures, we simply use a linear function of $b(T), T<T(\left\langle\Phi\right\rangle\approx 0.728)\approx 356~\text{MeV}$ which decreases monotonously with temperature and satisfies the condition of $b(0)=-8$ and $b(T\approx356~\text{MeV})\approx-15.02$. As shown in Figure~\ref{fig:b_T}, the absolute value of $b$ increases monotonously with temperature. We find that $b$ changes rather slowly in a very large interval of the temperature and sharply increases at $T\approx 710~\text{MeV}$. At this moment, we do not know how to understand such behavior of $b(T)$ on the physical level. One possible reason is that the thermal fluctuation should be included at very high temperatures, and thus this sharp increase in  $b$ will vanish with the inclusion of the fluctuation.

\begin{figure}[H]

	\includegraphics[scale=0.1]{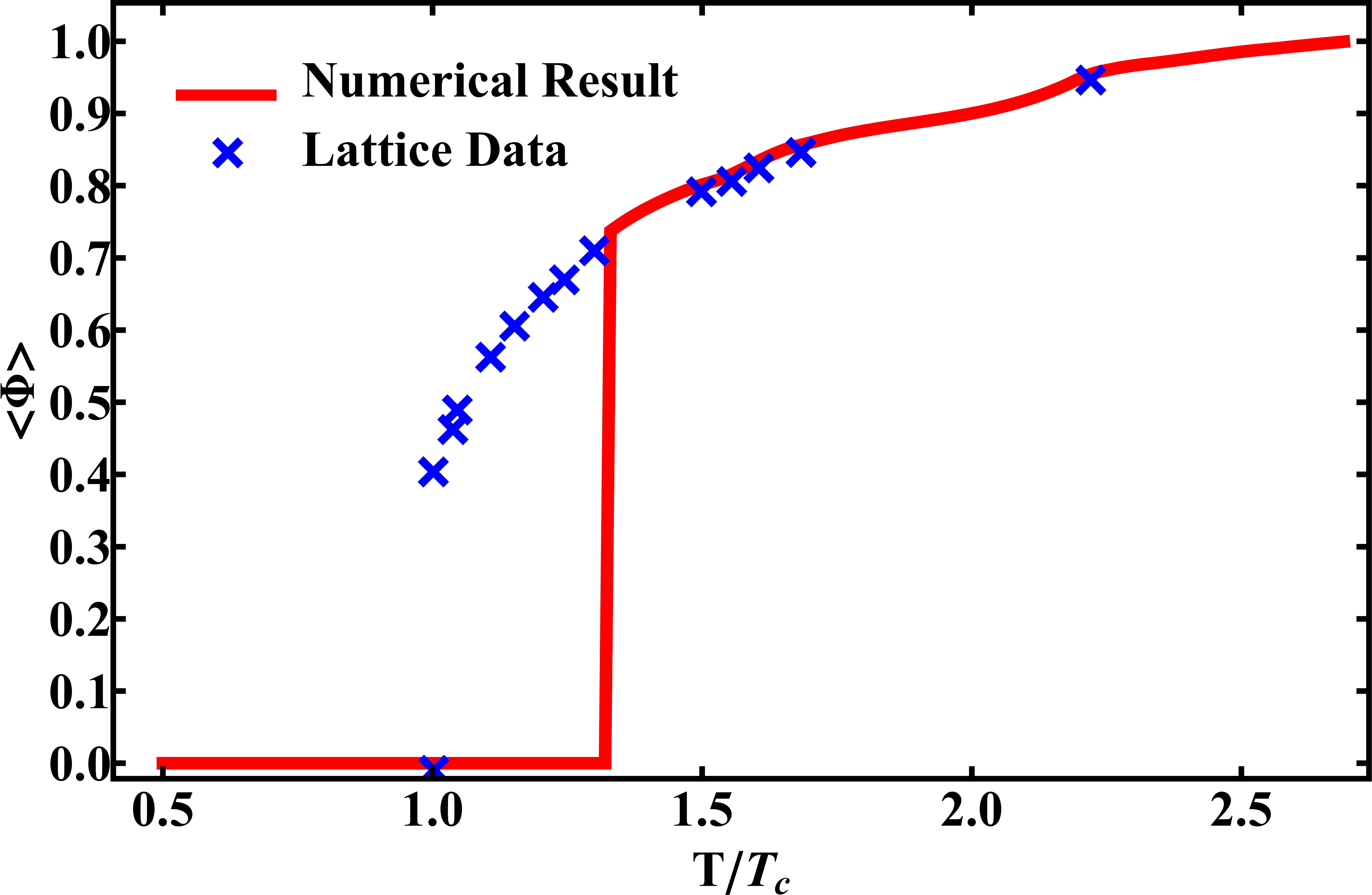}	
	\caption{\label{fig:Phi_T}The expectation value of the Polyakov loop as a function of temperature. Here, we normalize the variable $T$ by $T_{c}=270~\text{MeV}$ in order to compare the result with that of LQCD~\cite{Kaczmarek:2002mc,Ratti:2005jh}.}
\end{figure}

To show the reasonability of the fixed function $b(T)$, we evaluate the expectation value of the Polyakov loop as a function of $T$  with a comparison to the data of LQCD in Figure~\ref{fig:Phi_T}. It is obvious that the first-order phase transition is realized, and after the phase transition, we reproduce the result of LQCD qualitatively. We should note that the calculated transition temperature $\widetilde{T}_{c}=T(\left\langle\Phi\right\rangle\approx 0.728)\approx 356~\text{MeV} $ is larger than $270~\text{MeV}$. This enormous discrepancy is caused by the fact that when the phase transition happens, the minimum point of our Polyakov loop potential Equation~\eqref{eq:FinalPotent} will jump directly from zero to $\left\langle\Phi\right\rangle\approx 0.728$  so that we fit $b(T)$ starting from $\left\langle\Phi\right\rangle\approx 0.728$, and the data of LQCD from $\left\langle\Phi\right\rangle\approx0.4$ to $\left\langle\Phi\right\rangle\approx0.7$ cannot be used. The drastic jump of the minimum point may   result from MFA and might be remedied by the quantum corrections. Nevertheless, the primary goal of this work---setting up a possible relation between the trace anomaly of QCD and the color confinement and realizing the first-order phase transition---can be accomplished.

\begin{figure}[H]

	\includegraphics[scale=0.09]{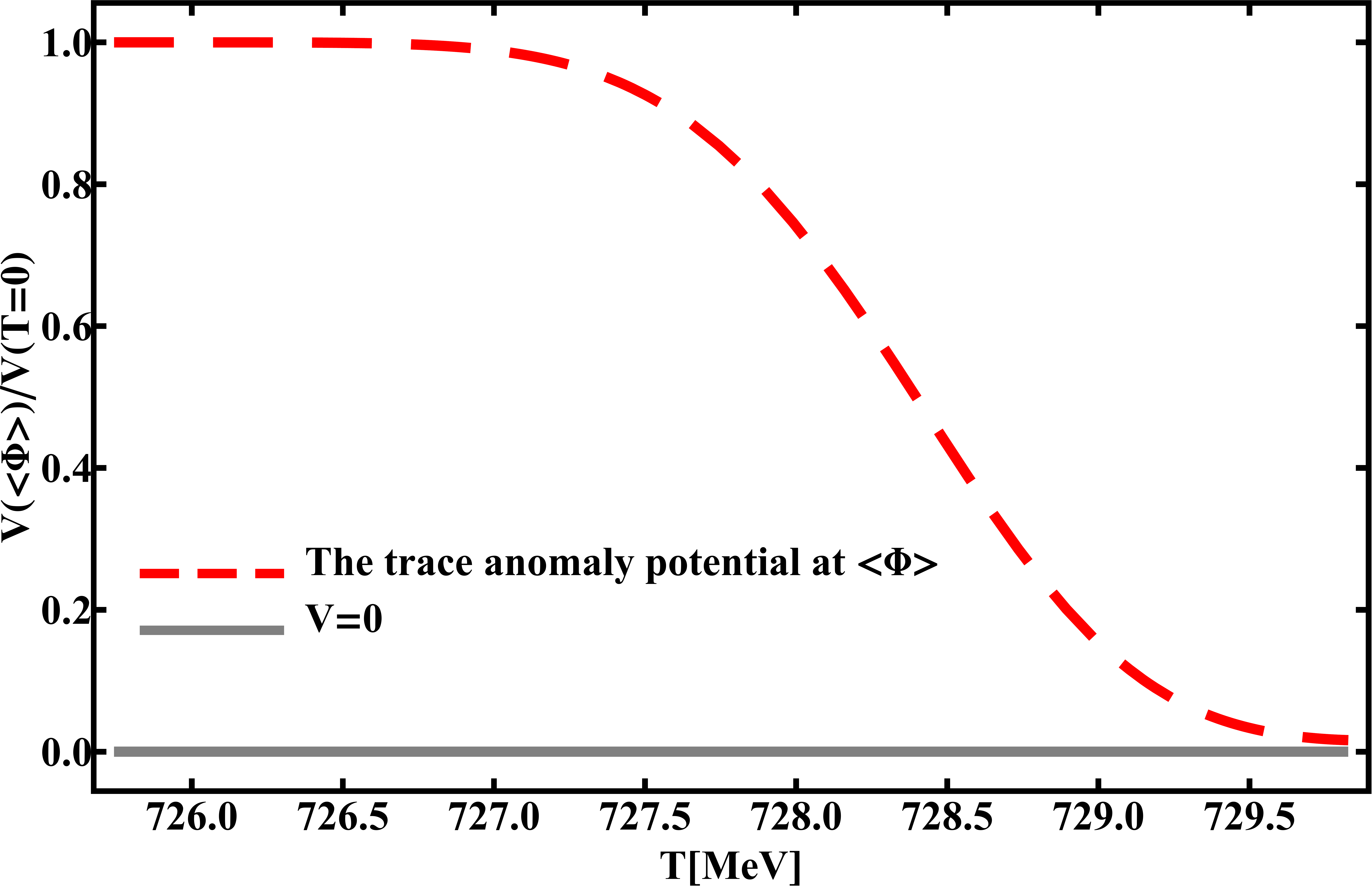}	
	\caption{\label{V_T}The value of Coleman--Weinberg type potential $V(\Phi,\Phi^\ast)$ at $\left\langle\Phi\right\rangle$ as a function of temperature.}
\end{figure}

We next check the trace anomaly in the thermal system by calculating the value of $V(\Phi=\left\langle\Phi\right\rangle)$, which is a function of temperature. The result presented in Figure~\ref{V_T} shows that, as expected, the trace anomaly potential at $\left\langle\Phi\right\rangle$ will be melted at a high-temperature limit and the scale symmetry will be restored. 

We finally want to emphasize that the behaviors of the above variables changing with the increase in  temperature are actually from the data of LQCD, and  the disappearance of the trace anomaly is especially the result of the constraint for the parameters, i.e., Equation~\eqref{eq:constraintabc}. In fact, we have no information about the parameters $a$, $b$, and $c$ as the functions of temperature without the help of LQCD, and we have not yet found the method that can calculate these parameters theoretically.  The holographic model of QCD is probably a possible way. Despite this, the first-order phase transition can be realized by the potential Equation~\eqref{eq:FinalPotent}, and this implies that the restoration of scale symmetry and the deconfinement phase transition may have a certain relationship.

\section{Trace Anomaly in Effective Models}\label{Tr ano EFT}
We next extend the above discussion on the pure gluon system to include quarks. In the presence of quarks, the deconfinement phase transition is no longer of the first order but a smooth crossover, as shown by LQCD~\cite{Ratti:2021ubw,Borsanyi:2010bp}. For this reason, it is of interest to investigate the deconfinement phase transition utilizing the low-energy effective quark models compensated by Polyakov potential. For simplicity, we will not consider the current quark mass term.

Firstly, we consider the $2+1$ flavor Polyakov--Nambu\textendash Jona--Lasino model (PNJL) model coupled to the background gluon field~\cite{Mattos:2021tmz}
\be
\mathcal{L}_{PNJL} & = & \bar{q}i\slashed{D}q+\dfrac{G_{s}}{2}\sum_{a=0}^{8}[(\bar{q}\lambda_{a}q)^2-(\bar{q}\gamma_{5}\lambda_{a}q)^2]   \nonumber\\
& & {} - \dfrac{G_{V}}{2}\sum_{a=0}^{8}[(\bar{q}\gamma_{\mu}\lambda_{a}q)^2+(\bar{q}\gamma_{\mu}\gamma_{5}\lambda_{a}q)^2] \nonumber \\
& & {} + K\{\text{det}_{f}[\bar{q}(1-\gamma_{5})q]+\text{det}_{f}[\bar{q}(1+\gamma_{5})q]\},
\label{eq:3fNJL}
\ee
where $D^{\mu}=\partial^{\mu}+i\delta_{0}^{\mu}g_{s}A^{0}_{a}\lambda_{a}/2$, with $g_{s}$ being the gauge coupling. It is clear that the Lagrangian~\eqref{eq:3fNJL} is not scale-invariant. The trace anomaly in the PNJL model is incorporated by guaranteeing the scale invariance of Equation~(\ref{eq:3fNJL}) and adding the Polyakov potential, which is constructed from Coleman--Weinberg-type potential~\eqref{eq:dilatonPotential}, which breaks the scale symmetry. Hence, the trace anomaly PNJL model, which we call  scale-PNJL (sPNJL), reads\vspace{12pt}
\be	\mathcal{L}_{sPNJL} & = & \bar{q}i\slashed{D}q + \dfrac{G_{s}}{2}\chi^{-2}\sum_{a=0}^{8}[(\bar{q}\lambda_{a}q)^2-(\bar{q}\gamma_{5}\lambda_{a}q)^2] \nonumber\\
& &{} - \dfrac{G_{V}}{2}\chi^{-2}\sum_{a=0}^{8}[(\bar{q}\gamma_{\mu}\lambda_{a}q)^2+(\bar{q}\gamma_{\mu}\gamma_{5}\lambda_{a}q)^2] \nonumber \\
& &{} + K\chi^{-5}\nonumber\\
& & {}\;\;\;\; \times \left\{\text{det}_{f}[\bar{q}(1-\gamma_{5})q]+\text{det}_{f}[\bar{q}(1+\gamma_{5})q]\right\} \nonumber\\
& &{} -\mathcal{U}(\Phi,\Phi^{*}),
\label{eq:TAPNJL}
\ee
with the Polyakov potential~\eqref{eq:FinalPotent}.

One may find that there are divergences in the interactions when the scale symmetry is restored, i.e., at $\chi=0$, if the couplings $G_{s}$, $G_{V}$, and $K$ are constants. This trouble may be overcome by assuming that these couplings are functions of $\chi$ and satisfy the following constraints:
\be
\lim_{\chi\to 0}G_{s}\chi^{-2}=\lim_{\chi\to 0}G_{V}\chi^{-2}=\lim_{\chi\to 0}K\chi^{-5}=0.
\label{eq:TAPNJLconstr}
\ee
This constraint implies that the interaction between quarks will be decreased in the deconfinement phase~\cite{Paeng:2011hy}. Furthermore, the constraints show that the sPNJL couplings are entangled with the Polyakov loop; that is, we can obtain the so-called entanglement-PNJL (EPNJL) model in Refs.~\cite{Sakai:2010rp,Sasaki:2011wu,DElia:2009bzj,deForcrand:2010he,Kogut:2004zg} from our sPNJL model.

Another model we want to analyze is the Polyakov--quark--meson model (PQMM)~\cite{Schaefer:2007pw,Schaefer:2009ui,Herbst:2010rf,Schaefer:2011ex}. The Lagrangian of the PQMM without the Polyakov potential reads~\cite{Mao:2009aq}
\be
\mathcal{L}_{PQMM}=\mathcal{L}_{quark}+\mathcal{L}_{meson},
\label{eq:PQMM}
\ee
where the $\mathrm{SU}(3)_{\text{R}}\times\mathrm{SU}(3)_{\text{L}}$ symmetric quark sector reads
\be
\mathcal{L}_{quark}=\bar{\psi}\left[i\slashed{D}-gT_{a}(\sigma_{a}+i\gamma_{5}\pi_{a})\right]\psi,
\label{eq:PQMMquark}
\ee
and the purely mesonic Lagrangian is
\be
\mathcal{L}_{meson} & = & \mathrm{Tr}\left(\partial_{\mu}\mathcal{M}^{\dagger}\partial^{\mu}\mathcal{M}\right) \nonumber\\
& & {} - \bm{\lambda}_{1}\left[\mathrm{Tr}(\mathcal{M}^{\dagger}\mathcal{M})\right]^2 - \bm{\lambda}_{2}\mathrm{Tr}\left[(\mathcal{M}^{\dagger}\mathcal{M})^2\right] \nonumber\\
& &{} +c\left[\mathrm{Det}(\mathcal{M})+\mathrm{Det}(\mathcal{M}^{\dagger})\right],
\label{eq:PQMMmeson}
\ee
with $\mathcal{M}$ being a complex $3\times 3$ matrix of meson fields.

In Equation~\eqref{eq:PQMMmeson}, $\bm{\lambda}_{1}$ and $\bm{\lambda}_{2}$ are the coupling constants of two possible quartic terms and $c$ is the coupling constant of the cubic term, which breaks $\mathrm{U(1)_{A}}$ symmetry. The details about the symmetries of PQMM can be found in Refs.~\cite{Mao:2009aq,Kawaguchi:2021nsa}.

One can easily check that in~ PQMM~\eqref{eq:PQMM}, only the last term in $\mathcal{L}_{meson}$ breaks scale symmetry. Therefore, the scale symmetry implemented model, which we call  scale-PQMM (sPQMM), becomes
\be
\mathcal{L}_{sPQMM}=\mathcal{L}_{quark}+\widetilde{\mathcal{L}}_{meson}-\mathcal{U}(\Phi,\Phi^{*}),
\label{eq:TAPQMM}
\ee
where the Polyakov potential~\eqref{eq:FinalPotent} is added and the scale symmetric meson sector is
\be
\widetilde{\mathcal{L}}_{meson}  & = & \mathrm{Tr}(\partial_{\mu}\mathcal{M}^{\dagger}\partial^{\mu}\mathcal{M}) \nonumber\\
& & {} - \bm{\lambda}_{1}\left[\mathrm{Tr}(\mathcal{M}^{\dagger}\mathcal{M})\right]^2-\bm{\lambda}_{2}\mathrm{Tr}\left[(\mathcal{M}^{\dagger}\mathcal{M})^2\right] \nonumber \\
& & {} + c\chi\left[\mathrm{Det}(\mathcal{M})+\mathrm{Det}(\mathcal{M}^{\dagger})\right].
\label{eq:SSPQMM}
\ee
The last term gives the coupling between the mesons and the dilaton. It is interesting to note that in the deconfined phase, $\chi=0$, the determinant terms in both PNJL and PQMM, which embody the  $\mathrm{U(1)_{A}}$ anomaly, are melted, anticipating the restoration of the $\mathrm{U(1)_{A}}$ symmetry at high temperatures. Since the large mass of the $\eta^\prime$ meson is generated by the $\mathrm{U(1)_{A}}$ anomaly, the decrease in its value will emerge in an extremely heated bath~\cite{Mei:2022dkd,Csorgo:2009pa,Benic:2012eu,Kwon:2012vb,Schaffner-Bielich:1999cux}.

To end this section, we make a brief comment on the mentioned sPNJL and sPQMM. The primary aim of this section is to demonstrate the method to construct the effective quark models in which the QCD trace anomaly is incorporated and what possible conclusions  result from  the models, for example, the derivation of EPNJL model and the description of the restoration of $\mathrm{U(1)_{A}}$ mentioned above. The numerical calculations based on these two low-energy effective quark models are not performed in this work, and precisely speaking, the current quark masses should be included when comparing the results with those of LQCD. Like the scenario used in Refs.~\cite{Ratti:2005jh,Roessner:2006xn}, one may use the Polyakov potential \eqref{eq:FinalPotent} with the parameters that we have fitted in terms of the LQCD data in pure {\textbf{glue}} sector and fix other parameters of the quark models with the input of some hadronic observables such as pion mass and its decay constant. However, more accurate parameter fixing may be performed when reproducing the crossover behavior of the deconfinement transition, and this issue will be investigated in our future works.

\section{Summary and Discussion}
We attempt to set up a relation between the trace anomaly of QCD and the deconfinement phase transition with respect to the observation that at extremely high temperatures and/or densities, the scale symmetry of QCD manifests and the deconfinement phase transition takes place. Considering that both the trace anomaly potential and the Polyakov loop potential have logarithmic forms and the conformal compensator and  the Polyakov loop are related by the configuration of the gluon field, we conjecture a relation between the expectation values of the conformal compensator $\chi$ and the Polyakov loop $\Phi$. 

It is found that the first-order phase transition in the pure glue sector can be realized phenomenologically by the Polyakov potential, which is composed of the trace anomaly potential $V$ and the quadratic term of $\Phi$. That is, the trace anomaly is locked to the deconfinemet~\cite{Liu:2023cse}. We should say that although the $\Phi^2$ is essential for realizing the first-order phase transition, we do not know what its origin is. This issue may imply that the Polyakov potential includes other effects of the strongly interacting gluons except for the trace anomaly, and the kinetic energy of gluons should be determined~\cite{Fukushima:2003fw}.

A fly in the ointment is that the critical temperature evaluated by our Polyakov potential is larger than that of LQCD; this evident discrepancy may be caused by the MFA used in this work. Actually, near the critical temperature, the thermal fluctuation will impact the properties of the system, and we will look for a method to include the thermal fluctuation in   future works. Nevertheless, the primary purpose of this work is to illustrate the possible relationship between the restoration of scale symmetry and the deconfinement phase transition at extreme conditions and to manifest the first-order phase transition by our Polyakov potential. We indeed accomplished this purpose.

Since the quark degrees of freedom should be included to investigate the deconfinement phase transition at high temperatures, we constructed the Lagrangians of sPNJL and sPQMM in the chiral limit in which the Polyakov potential based on the trace anomaly potential is incorporated. On the one hand, we find that the conformal compensator $\chi$ has the power to build a bridge between the restoration of $\mathrm{U(1)_{A}}$ symmetry and the deconfinement phase transition, and on the other hand, it is interesting that the previously proposed EPNJL model can be obtained naturally from our models.

In this work, we  are just trying to conjecture a possible relationship between the restoration of scale symmetry and the deconfinement transition at high temperatures based on a rather phenomenological ansatz. Hence, there is a very fundamental question: How can we build a bridge theoretically between the trace anomaly and the deconfinement phase transition based on some low-energy effective models or theories of QCD? This will be clarified in future works.
\vspace{6pt} 



\authorcontributions{Both authors contributed equally to this article. Both authors have read and agreed to the published version of the manuscript. 
}

\funding{The work of Y. L. M. was supported in part by the National Science Foundation of China
	(NSFC) under Grant No. 12347103, No. 11875147 and No. 12147103. 
}

\dataavailability{No new data were created or analyzed in this study. 
}




\acknowledgments{We would like to thank K. F. Liu and Y. L. Wu for their valuable discussions and suggestions. The work of Y.~L. M. was supported in part by the National Science Foundation of China (NSFC) under Grant No. 12347103, No. 11875147 and No. 12147103. 
}

\conflictsofinterest{The authors declare no conflicts of interest. 
}



\begin{adjustwidth}{-\extralength}{0cm}

\reftitle{References}

\PublishersNote{}
\end{adjustwidth}
\end{document}